\documentclass[intlimits,twoside,a4paper]{article}

\usepackage[T2A]{fontenc} %T2A
\usepackage[utf8]{inputenc} %cp1251
\usepackage[greek,ukrainian,english]{babel}

\usepackage[eqsecnum]{cmpj3}
\usepackage{multirow}
\usepackage{graphicx}
\usepackage{longtable}

\issue{2026}{29}{1}{13801}
\doinumber{10.5488/CMP.29.13801}

\title[Modelling amino acid hydration]%
{The application of Kirkwood-Buff theory to study hydration properties of $\alpha$-amino acids%
\thanks{Dedicated to the 75th birthday of Prof. Stefan Soko\l owski.}}
\author[Z. \v{S}tefani\v{c}, B. Hribar-Lee]{Z. \v{S}tefani\v{c}\orcid{0009-0008-3837-3941}, B. Hribar-Lee\orcid{0000-0002-9029-588X}}
\address{
 Faculty of Chemistry and Chemical Technology, University of Ljubljana, Ve\v{c}na pot 113, SI-1000 Ljubljana, Slovenia
}


\date{Received 20 November 2025; revised 29 December 2025; accepted 29 December 2025; published~30~March~2026}

\begin{document}

\maketitle

\begin{abstract}
Protein conformational stability and function depend on non-covalent interactions that are strongly influenced by the surrounding environment. To explore protein properties, amino acids are often utilized as model systems. In this study, we determined the densities of seven $\alpha$-amino acids in aqueous solutions between 278.15 K and 308.15 K and calculated the apparent molar volumes. Linear extrapolation yielded standard molar volumes, which were analyzed to characterize amino-acid hydration. The contributions of side chains to the standard molar volume were determined relative to glycine. The standard molar volume increased with temperature, indicating reduced electrostriction of water around the amino acids, consistent with lower hydration numbers at higher temperatures. We employed the Ornstein–Zernike integral equation with hypernetted-chain closure and a coarse-grained Lennard-Jones bead model to calculate pair correlation functions and Kirkwood–Buff integrals, from which standard molar volumes were obtained. The model reproduced the experimental standard molar volumes very well.
%
%
%\keywords Up to six keywords (\href{https://physh.aps.org/browse}{Physics Subject Headings})
\printkeywords
%
%\pacs Up to six PACS numbers (optional)
\end{abstract}

\section{Introduction}

Proteins are important macromolecules within biological systems, playing essential roles in various physiological processes. Their conformational stability, and therefore their functionality, depends on non-covalent interactions such as electrostatic, hydrophobic, and hydrogen bonds \cite{Yan04,Panuszko19,Bagchi13}. These interactions are strongly influenced by the surrounding environment, especially solvent. The complexity of proteins poses significant challenges to direct thermodynamic studies. Therefore, to understand these interactions on protein behaviour, we study model compounds such as amino acids and peptides, which are the building blocks of proteins \cite{Zhao06,Lolicoeur86,Mota14}.

Remarkable scientific work has been done on the thermodynamics of amino acids in aqueous electrolyte solutions \cite{Yan04,Zhao06,Lolicoeur86,Mota14,Banipal12,Riyazuddeen12,Mallick06,Rajagopal12,Iulian13,Romero18,Rodriguez17}. These studies used the partial molar volumes of transfer to assess interactions between ions and zwitterionic centres of amino acids. Investigation of hydration numbers allowed for the quantitative explanations of dehydration of amino acids with temperature and salt concentration \cite{Yan04,Lolicoeur86,Mota14}. However, to the best of our knowledge, no studies have examined and compared all different types of amino acid side chains and how they influence the hydration of amino acids in pure water.

To study the microstructure of solutions that would allow the molecular interpretation of experimental data, researchers usually resort to computer simulations \cite{Lumb1987,Sokolowski2017}. Computer simulation data provide full information on the various pair correlation functions that measure the extent of correlation between pairs of molecules. To compare these data with experimental ones, Kirkwood-Buff (KB) theory of solutions, which uses averages of the pair correlation functions, is often used to analyze computer simulation results \cite{Vergara2002}. As such, KB theory represents a valuable tool for investigation of the changes in the solvation microstructure of solutions, as well as for description of thermodynamic properties of solutions in an exact
manner over the whole concentration range \cite{Zielkiewicz1998}. One of the advantages of KB theory is that it not only allows the analysis of computer simulation results, but can also use pair correlation functions obtained from other statistical thermodynamic theories, such as Ornstein-Zernike (OZ) integral equation~\cite{Sokolowski2}, which are not subject to relatively large statistical errors present in computer simulations of complex systems. KB theory was successfully applied to study the thermodynamics of solution with small deviations from ideal behavior \cite{Matteoli1997}, as well as more complex solutions, such as biological systems~\cite{Pierce2008}. Hydration properties, in particular, were studied in \cite{Imai00,Shulgin05,Simon22}. 

In this work, we wanted to test the applicability of KB theory in combination with the OZ integral equation, and a simple coarse-grained amino acid model to study the volumetric and hydration properties of seven proteinogenic amino acids of various types: arginine, aspartic acid, glutamic acid, glycine, lysine, serine, and tryptophan. These amino acids were selected to encompass different side-chain group types. We examine how volumetric properties of these amino acids, which are a consequence of their hydration, are influenced by different side chains. Subsequently, we employ newly obtained experimental data to calculate the standard molar volumes of amino acids as well as their hydration numbers, providing insights into their hydration properties relevant to proteins \cite{Yan04,Zhao06,Millero78}. The standard molar volumes of these amino acids were then estimated using KB theory, for which we used pair correlation functions obtained from the OZ integral equation theory in combination with a simple coarse-grained amino acid model.

The paper is organized as follows: we first outline our methodology for sample preparation, experimental procedures and the theoretical approach. Subsequently, we present and discuss our findings regarding the volumetric properties of the studied amino acids. Conclusions are given at the end.

\section{Materials and methods}
\subsection{Experimental}

L-arginine (Sigma-Aldrich, reagent grade, $\geqslant 98\%$ purity), and L-lysine [Sigma-Aldrich, $\geqslant 98\% $ purity (TLC)] were used after being dried over P$_{2}$O$_{5}$ in a vacuum desiccator for more than 72 hr before use. The moisture content was determined using Karl Fisher titration (Mettler Toledo DL38). L-aspartic acid [Sigma-Aldrich, R.G., $\geqslant 98\%$ purity (HPLC)], L-glycine (Sigma-Aldrich, GR), L-glutamic acid [Sigma-Aldrich, R. G., $\geqslant 99\%$ purity (HPLC)], L-serine [Sigma-Aldrich, ReagentPlus, $\geqslant 99\%$ purity (HPLC)], L-tryptophan [Sigma-Aldrich, R.G., $\geqslant 98\%$ purity (HPLC)] were used without further purification.

Solutions were prepared using deionized water (Rephile, Purist UV, with conductivity of 0.054~\textmu{}S at~25\textcelsius). The weights were determined on a Mettler Toledo MS304TS/M00 balance, which had a precision of 0.1 mg. All solutions were prepared by molarity from stock solutions of amino acids. Samples were stirred for three hours to achieve homogenous solutions. Due to the hydrophobic nature of tryptophan, an ultrasonic bath [Elma, Elmasonic S 40 (H)] was used to help with the dissolution, followed by the measurement of molarity using NanoDrop One (Thermo Fisher Scientific). After preparation, and just before measurement, solutions were degassed for 30 minutes (TA instruments, Degassing Station). 

The densities of the solutions were measured with a density and sound velocity meter: DSA 5000 M (Anton Paar) with an accuracy of $\pm0.7 \times 10^{-5}$~g/cm$^3$ (according to the manufacturer). The measuring cell was calibrated with dry air and ultra pure water at atmospheric pressure. The cell is temperature-controlled by a built-in Peltier thermostat with a thermal stability of $\pm$0.001 K. An average of two measurements was considered for each sample.

\subsection{Theory and model}

To complement the density measurements and obtain a microscopic picture of hydration, we modelled each amino-acid solution as a binary mixture of water (1) and a single amino acid (2), where both species are represented by one isotropic Lennard–Jones (LJ) bead. Pair interactions were defined as
\begin{equation}
  u_{ij}(r) = 4\varepsilon_{ij}\left[\left(\frac{\sigma_{ij}}{r}\right)^{12}
  - \left(\frac{\sigma_{ij}}{r}\right)^{6}\right],
\end{equation}
with the Lorentz–Berthelot mixing rules for the cross parameters,
\begin{equation}
  \sigma_{12} = \frac{\sigma_{11} + \sigma_{22}}{2}, \qquad
  \varepsilon_{12} = \sqrt{\varepsilon_{11}\varepsilon_{22}}.
\end{equation}

Water was treated as a single LJ site chosen to reproduce its short-range structure in the SPC/E model ($\sigma_{11}$ = 0.316 nm and $\varepsilon_{11}/k_\mathrm{\text{B}}T$ = 0.26) \cite{Berendsen87}, while the amino acids were assigned van der Waals-based bead sizes $\sigma_{22}$ and interaction strengths $\varepsilon_{22}/k_\mathrm{\text{B}}T$ (table~\ref{tab:lj-aa}) \cite{Blanco13}. We solved the Ornstein–Zernike (OZ) equation for the binary mixture using the hypernetted-chain (HNC) closure,

\begin{equation}
  h_{ij}(r) = \exp\!\big[-\beta u_{ij}(r) + \gamma_{ij}(r)\big] - 1 - \gamma_{ij}(r),
  \qquad
  \gamma_{ij}(r) = h_{ij}(r) - c_{ij}(r),
\end{equation}
where $h_{ij}(r)$ and $c_{ij}(r)$ are the total and direct correlation functions, respectively, and $\beta = 1/(k_\mathrm{\text{B}}T)$.

The amino acids were treated in the tracer limit (effectively infinite dilution), which is numerically robust and directly relevant to the standard (infinite-dilution) molar volume reported below.

\begin{table}[t]
  \centering
  \caption{Lennard--Jones parameters for the amino-acid beads:
  $\sigma_{22}$ (nm) and $\varepsilon_{22}/k_\mathrm{\text{B}}T$ at $T = 298.15$~K, taken from~\cite{Blanco13}.}
  \label{tab:lj-aa}
  \begin{tabular}{lcc}
    \hline
    Amino acid & $\sigma_{22}$ (nm) & $\varepsilon_{22}/k_\mathrm{\text{B}}T$ \\
    \hline
    Gly & 0.431 & 1.05 \\
    Ser & 0.528 & 0.99 \\
    Asp & 0.583 & 0.94 \\
    Glu & 0.640 & 0.93 \\
    Trp & 0.670 & 1.52 \\
    Arg & 0.732 & 1.01 \\
    Lys & 0.703 & 0.88 \\
    \hline
  \end{tabular}
\end{table}

From the converged pair distribution functions,
\begin{equation}
  g_{ij}(r) = 1 + h_{ij}(r),
\end{equation}
we computed the Kirkwood--Buff (KB) integrals,
\begin{equation}
  G_{ij} = 4\piup \int_0^\infty [g_{ij}(r) - 1]\, r^2 \,\mathrm{d}r,
\end{equation}
which quantify the local accumulation or depletion of species $j$ around a tagged particle of species $i$. In the infinite-dilution limit, the KB framework connects these integrals to partial molar quantities. In particular, the standard molar volume of the solute at infinite dilution is
\begin{equation}
  V^{\circ}_{\phi} = N_\mathrm{A}\left(\frac{1}{\rho_1} + G_{11} - G_{12}\right),
\end{equation}
where $\rho_1$ is the number density of water molecules and $N_\mathrm{A}$ is Avogadro’s number.

All OZ--HNC calculations were performed at $T = 298.15$~K to match the main experimental dataset, and the same seven amino acids as in the density study (Arg, Asp, Glu, Gly, Lys, Ser, and Trp) were considered.

\section{Results and discussion}

\begin{figure}[hbt]
\centerline{\includegraphics[scale=0.3]{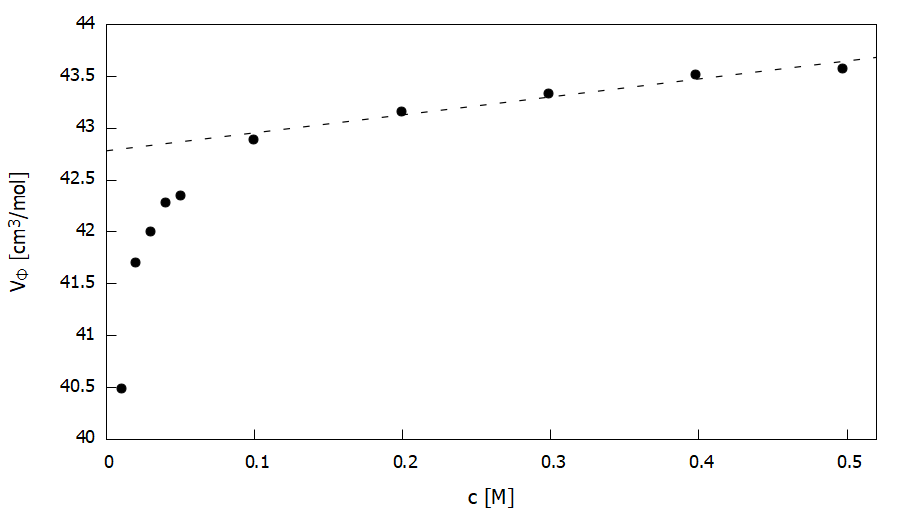}}
\caption{The apparent molar volumes ($V_{\phi}$) of L-glycine at 298.15 K across various molar concentrations (the size of the symbol correspond to the estimated measurement error). The dotted line represents the extrapolation to apparent molar volumes at infinite dilution ($V^{\circ}_{\phi}$). Note that the discrepancies from linearity at low amino acid concentrations are due to a low density measurement accuracy of the densimeter at low concentration range.}
\label{fig:1}
\end{figure}

The measured densities (g/cm$^{3}$) (found in table \ref{tab:3}) of aqueous solutions of amino acids were used to calculate the apparent molar volumes $V_{\phi}$ (cm$^{3}$/mol) as:
\begin{equation} \label{eq:1}
V_{\phi} = \frac{M}{\rho_{\circ}} - \frac{1000 (1 - \frac{\rho}{\rho_{\circ}})}{c},
\end{equation} 
where $c$ is the molar concentration of the solution, $M$ is the molar mass of the amino acid, and $\rho$ and $\rho_{\circ}$ are the densities of the solutions and solute, respectively. The determined apparent molar volume data were well represented with a linear equation:
\begin{equation} \label{eq:2}
V_{\phi} = V^{\circ}_{\phi} + S_{v}c,
\end{equation}
where $V_{\phi}^{\circ}$ is the apparent molar volume at infinite dilution and is therefore the same as the standard molar volume. $S_{v}$ is the experimental slope and it indicates the strength of the solute-solute interactions \cite{Zhao06}. The linear representation of the obtained partial molar volumes of glycine at temperature of 298.15 K, and the extrapolation to the infinite dilution are shown in figure \ref{fig:1}. The obtained standard molar volumes and experimental slopes are for all the amino acids studied summarized in table \ref{tab:Vphi_Sv} along with their standard deviations.
\begin{table}[!hbtp]
\caption{Standard molar volumes ($V^{\circ}_{\phi}$) and volumetric
pairwise interactions coefficient ($S_{v}$) for $\alpha$-amino acids in water
at various temperatures (values in parentheses are estimated uncertainties).}
\label{tab:Vphi_Sv}
\centering
\small
\setlength{\tabcolsep}{6pt}
\renewcommand{\arraystretch}{1.15}

% --- PRVI DEL: 178.15 K in 288.15 K ---
\begin{tabular}{lcccc}
\hline
\multirow{2}{*}{Amino acid} &
\multicolumn{2}{c}{178.15 K} &
\multicolumn{2}{c}{288.15 K} \\
 & $V^{\circ}_{\phi}$ [cm$^{3}$\,mol$^{-1}$] & $S_v$ &
   $V^{\circ}_{\phi}$ [cm$^{3}$\,mol$^{-1}$] & $S_v$ \\
\hline
Lysine        & 106.51 (0.03) & 0.77  (0.2)  & 107.98 (0.03) & 0.78  (0.2)  \\
Arginine      & 120.98 (0.09) & 2.21  (0.1)  & 122.76 (0.09) & 1.96  (0.09) \\
Aspartic acid &  67.34 (0.3)  & 99.04 (1.3)  &  68.66 (0.3)  & 116.45 (1.6) \\
Glycine       &  40.70 (0.06) &  1.98 (0.1)  &  41.89 (0.06) &  1.84 (0.2)  \\
Serine        &  58.02 (0.1)  &  2.51 (0.4)  &  59.33 (0.1)  &  2.46 (0.4)  \\
Glutamic acid &  82.33 (0.3)  & 55.93 (9.3)  &  84.07 (0.3)  & 61.65 (10.1) \\
Tryptophan    & 141.74 (0.7)  &192.71 (19.9)  & 142.98 (0.7)  &224.10 (22.0)  \\
\hline
\end{tabular}

\vspace{0.5em} % ~ ena do dve vrstici prostora, po potrebi spremeni

% --- DRUGI DEL: 298.15 K in 308.15 K ---
\begin{tabular}{lcccc}
\hline
\multirow{2}{*}{Amino acid} &
\multicolumn{2}{c}{298.15 K} &
\multicolumn{2}{c}{308.15 K} \\
 & $V^{\circ}_{\phi}$ [cm$^{3}$\,mol$^{-1}$] & $S_v$ &
   $V^{\circ}_{\phi}$ [cm$^{3}$\,mol$^{-1}$] & $S_v$ \\
\hline
Lysine        & 109.16 (0.03) & 0.61  (0.1)  & 110.05 (0.03) & 0.68  (0.1)  \\
Arginine      & 124.20 (0.09) & 1.70  (0.2)  & 125.31 (0.10) & 1.70  (0.1)  \\
Aspartic acid &  70.54 (0.3)  &102.96 (1.34)  &  71.36 (0.3)  &114.19 (1.5)  \\
Glycine       &  42.78 (0.07) &  1.73 (0.2)  &  43.44 (0.06) &  1.67 (0.2)  \\
Serine        &  60.41 (0.1)  &  2.22 (0.4)  &  61.19 (0.1)  &  2.24 (0.4)  \\
Glutamic acid &  85.66 (0.3)  & 63.52 (10.1)  &  86.83 (0.3)  & 65.43 (10.6)  \\
Tryptophan    & 145.12 (0.7)  &213.31 (21.0)  & 146.63 (0.7)  &219.57 (22.0)  \\
\hline
\end{tabular}

\end{table}

The values of standard molar volumes for ionizable amino acids (glutamic acid, aspartic acid, lysine, arginine, and tryptophane) are reported as measured in aqueous solutions. Contributions of ionization on density measurement were considered negligible with concentrations used in the study. All amino acids were presumed to be in their zwitterionic form.

Density measurements at lower concentrations consistently showed poor linear agreement with measurements at higher concentrations (shown in figure \ref{fig:1}). We believe that these measurements do not coincide due to low density measurement accuracy of the densimeter, since the change in concentration was not well reflected in the change of the density measurement. Thus, the accuracy of the densimeter used limits this method of standard molar volume calculation at lower concentrations, while the solubility of a the given amino acid constrains it at higher concentrations. 

\begin{figure}[htb]
\centerline{\includegraphics[scale=0.3]{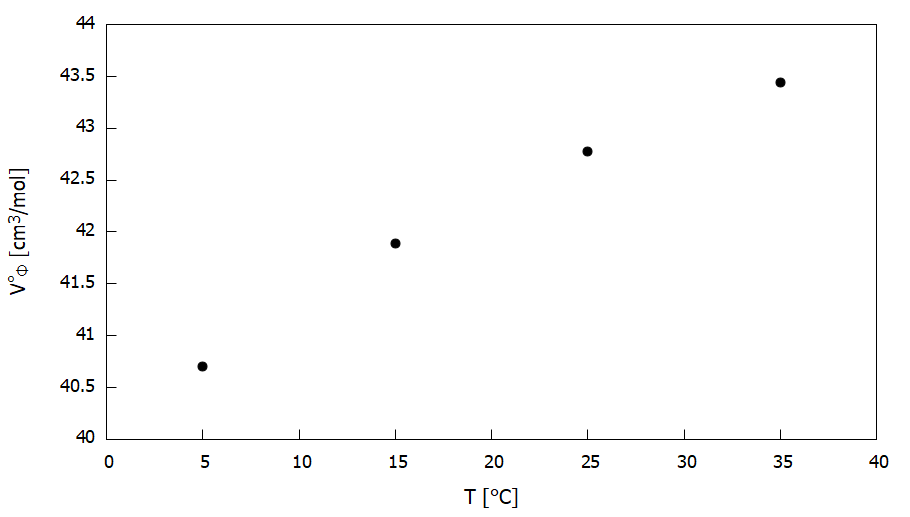}}
\caption{The partial molar volumes at infinite dilution ($V^{\circ}_{\phi}$) of L-glycine at four different temperatures: 278.15 K, 288.15 K, 298.15 K, and 308.15 K (the size of the symbol 
	correspond to the estimated measurement error).}
\label{fig:2}
\end{figure}

Standard molar volume values of amino acids increase linearly with temperature (for glycine shown in figure \ref{fig:2}), which can be explained with the decreased electrostriction of water around the amino acid molecules at higher temperatures \cite{Mota14}. Temperature dependence of standard molar volume provides valuable information about solute-solvent interactions since the solute-solute interactions can be neglected at infinite dilution.

To check the quality of our data we compared them to the data collected by Zhao \cite{Zhao06} at different concentrations. Results obtained in our study were in good agreement with previously available data, with the largest deviation found for aspartic acid at temperature of 308.15 K, where the value reported was 75.84 cm$^{3}$\,mol$^{-1}$ (ours was 71.36 cm$^{3}$\,mol$^{-1}$). This deviation could be caused by the reactivity of aspartic acid in aqueous solutions, where it undergoes isomerization, making repeatability of result harder \cite{Nakayoshi21}. 

Side chain contributions to standard molar volume were calculated as the difference in the quantity between each amino acid and glycine \cite{Lolicoeur86}:
\begin{equation} \label{eq:3}
V_{\phi}^{\circ} = V_{\phi}^{\circ}(\text{amino acid}) - V_{\phi}^{\circ}(\text{glycine}).
\end{equation} 

These contributions were consistent for all measured temperatures and increased with the molar weight of the side chains. The only exceptions were glutamic acid and lysine, with the latter having a slightly lower molar weight (146.19 g/mol) than the former (147.1 g/mol) and yet producing a significant rise in side-chain contribution. A possible explanation is that the amino group coupled with a longer alkyl chain of lysine contributes more to the molar volume of the amino acid than the carboxyl group of glutamate. 

Results for the standard molar volumes were also interpreted based on a continuum model of the solution using the Millero method \cite{Millero78,Millero64}. This model uses density data for the calculation of the hydration numbers of nonelectrolytes, such as amino acids. The hydration number, $n_{\text{H}}$, reveals the hydration degree of a solute in water. According to the model, the hydration number can be calculated via this equation:
\begin{equation} \label{eq:4}
n_{\text{H}} = \frac{V^{\circ}_{\phi} - V^{\circ}_{\phi, \text{int}}}{V^{\circ}_{\text{e}} - V^{\circ}_{\text{b}}},
\end{equation} 
where $V_{\text{e}}^{\circ}$ is the partial molar volume of electrostricted water and $V_{\text{b}}^{\circ}$ is the molar volume of the bulk water. These can be calculated using the procedure described by Millero \cite{Millero64} and can be found in table~\ref{tab:2}. $V_{\phi, \text{int}}^{\circ}$ is the intrinsic volume of the solute molecule and is calculated via equation (\ref{eq:5}) as proposed by Millero~\cite{Millero64}.

\begin{equation} \label{eq:5}
V^{\circ}_{\phi, \text{int}} = \left( \frac{0.7}{0.634}\right) V^{0}_{\phi, \text{cryst}}\,.
\end{equation} 
Here, $V_{\phi, \text{cryst}}^{\circ}$ is the crystal molar volume, calculated using equation
%~(\ref{eq:6}), where $\rho_{\text{cryst}}$ denotes the crystalline density taken from the study of Berlin and Pallansh~\cite{Berlin68}.

\begin{equation} \label{eq:6}
V^{\circ}_{\phi, \text{cryst}} = \frac{M}{\rho_{\text{cryst}}},
\end{equation} 
where $\rho_{\text{cryst}}$ denotes the crystalline density taken from the study of Berlin and Pallansh~\cite{Berlin68}.
Hydration numbers of amino acids are given in table~\ref{tab:2}. One can see that, as expected, the hydration number increases with the size of the amino acid and decreases with the temperature. This indicates that the temperature dependence of standard molar volume is determined by the influence of temperature on electrostriction of water by the zwitterionic centres of amino acids.

\begin{table}[!hbtp]
\caption{Hydration numbers, $n_{\text{H}}$, of amino acids in aqueous solutions at various temperatures.}
\vspace{2mm}
\label{tab:2}
\centering
\footnotesize
\begin{tabular}{l c c c c }
\hline
Amino acid & 278.15 K & 288.15 K & 298.15 K & 308.15 K \\
\hline
\mbox{$V^{\circ}_{\text{e}} - V^{\circ}_{\text{b}}$} [cm$^3$/mol] & $-2.6$ & $-2.9$ & $-3.3$ & $-4.0$ \\
\hline
Glycine & 4.3 & 3.4 & 2.8 & 2.1 \\
Serine & 5.9 & 4.8 & 3.9 & 3.0 \\
Aspartic acid & 8.6 & 7.3 & 5.8 & 4.6 \\
Glutamic acid & 8.2 & 6.8 & 5.4 & 4.3 \\
Lysine & 9.2 & 7.8 & 6.5 & 5.1 \\
Arginine & 9.4 & 7.9 & 6.5 & 5.1 \\
Tryptophan & 12.0 & 10.4 & 8.5 & 6.6 \\
\hline
\end{tabular}
\end{table}

\begin{table}[!hbtp]
\caption{Density measurements ($\rho$) and apparent molar volume ($V_{\phi}$) 
for different amino acids.}

\label{tab:3}
\centering
\footnotesize
\setlength{\tabcolsep}{4pt}   % po potrebi povečaj / zmanjšaj širino stolpcev
\renewcommand{\arraystretch}{1.1}

\begin{tabular}{c c c c c c c c c}
\hline
 & \multicolumn{2}{c}{278.15 K} & \multicolumn{2}{c}{288.15 K} & 
   \multicolumn{2}{c}{298.15 K} & \multicolumn{2}{c}{308.15 K} \\
\hline
c [M] & $\rho$ & $V_{\phi}$ & $\rho$ & $V_{\phi}$ & $\rho$ & $V_{\phi}$ & $\rho$ & $V_{\phi}$ \\
 & [g/cm$^{3}$] & [cm$^{3}$/mol] & [g/cm$^{3}$] & [cm$^{3}$/mol] &
   [g/cm$^{3}$] & [cm$^{3}$/mol] & [g/cm$^{3}$] & [cm$^{3}$/mol] \\
\hline

\multicolumn{9}{l}{Glycine}\\
0     & 0.9999638 &        & 0.9990996 &        & 0.997047  &        & 0.9940319 &        \\
0.10  & 1.003366  & 40.86  & 1.002389  & 42.03  & 1.000259  & 42.89  & 0.997192  & 43.55  \\
0.20  & 1.006715  & 41.12  & 1.005626  & 42.29  & 1.003417  & 43.17  & 1.000302  & 43.80  \\
0.30  & 1.010031  & 41.32  & 1.008834  & 42.48  & 1.006552  & 43.33  & 1.003388  & 43.97  \\
0.40  & 1.013299  & 41.54  & 1.012000  & 42.67  & 1.009646  & 43.52  & 1.006435  & 44.15  \\
0.50  & 1.016590  & 41.63  & 1.015188  & 42.75  & 1.012768  & 43.58  & 1.009508  & 44.20  \\
\hline

\multicolumn{9}{l}{Serine}\\
0.04  & 1.001845  & 57.98  & 1.000932  & 59.25  & 0.998841  & 60.34  & 0.995803  & 61.10  \\
0.05  & 1.002312  & 58.04  & 1.001386  & 59.33  & 0.999286  & 60.41  & 0.996241  & 61.19  \\
0.10  & 1.004633  & 58.32  & 1.003643  & 59.63  & 1.001496  & 60.70  & 0.998421  & 61.49  \\
0.20  & 1.009224  & 58.71  & 1.008107  & 60.03  & 1.005874  & 61.06  & 1.002738  & 61.85  \\
0.30  & 1.013776  & 58.97  & 1.012538  & 60.27  & 1.010222  & 61.28  & 1.007027  & 62.07  \\
0.40  & 1.018348  & 59.05  & 1.016993  & 60.33  & 1.014596  & 61.32  & 1.011343  & 62.11  \\
0.50  & 1.022929  & 59.08  & 1.021461  & 60.34  & 1.018985  & 61.32  & 1.015674  & 62.10  \\
\hline

\multicolumn{9}{l}{Aspartic acid}\\
0.001 & 1.000045  & 51.46  & 0.999184  & 48.28  & 0.997129  & 50.80  & 0.994115  & 49.84  \\
0.004 & 1.000236  & 64.68  & 0.999369  & 65.44  & 0.997309  & 67.44  & 0.994294  & 67.62  \\
0.007 & 1.000215  & 97.02  & 0.999347  & 97.65  & 0.997285  & 99.21  & 0.994266  &100.07  \\
0.010 & 1.000610  & 68.13  & 0.999732  & 69.58  & 0.997663  & 71.37  & 0.994641  & 72.29  \\
0.013 & 1.000798  & 68.58  & 0.999915  & 70.10  & 0.997843  & 71.75  & 0.994817  & 72.81  \\
0.016 & 1.000986  & 68.87  & 1.000097  & 70.49  & 0.998019  & 72.23  & 0.994994  & 73.08  \\
0.019 & 1.001165  & 69.54  & 1.000269  & 71.28  & 0.998190  & 72.83  & 0.995160  & 73.84  \\
0.022 & 1.001351  & 69.70  & 1.000450  & 71.45  & 0.998368  & 72.94  & 0.995332  & 74.12  \\
0.025 & 1.001538  & 69.79  & 1.000631  & 71.57  & 0.998544  & 73.11  & 0.995506  & 74.26  \\
0.030 & 1.001845  & 70.05  & 1.000930  & 71.82  & 0.998836  & 73.36  & 0.995795  & 74.45  \\
\hline

\multicolumn{9}{l}{Glutamic acid}\\
0.010 & 0.999047  &238.92  & 0.998197  &237.71  & 0.996145  &238.14  & 0.993126  &239.25  \\
0.015 & 0.999050  &208.13  & 0.998199  &207.43  & 0.996144  &208.01  & 0.993124  &208.97  \\
0.020 & 1.000480  &121.29  & 0.999604  &121.99  & 0.997538  &122.91  & 0.994518  &123.53  \\
0.025 & 1.000133  &140.36  & 0.999262  &140.75  & 0.997197  &141.54  & 0.994174  &142.29  \\
0.030 & 1.000530  &128.24  & 0.999644  &129.08  & 0.997572  &129.99  & 0.994544  &130.82  \\
0.035 & 1.000419  &134.11  & 0.999533  &134.85  & 0.997455  &135.86  & 0.994421  &136.82  \\
0.040 & 1.001508  &108.48  & 1.000598  &109.73  & 0.998510  &110.84  & 0.995471  &111.78  \\
0.050 & 1.001813  &110.11  & 1.000894  &111.30  & 0.998796  &112.44  & 0.995754  &113.32  \\
\hline

\multicolumn{9}{l}{Lysine}\\
0.10  & 1.001956  &106.58  & 1.001025  &108.06  & 0.998923  &109.21  & 0.995882  &110.12  \\
0.15  & 1.005897  &106.65  & 1.004826  &108.12  & 1.002634  &109.28  & 0.999535  &110.17  \\
0.20  & 1.007878  &106.64  & 1.006738  &108.11  & 1.004500  &109.26  & 1.001372  &110.16  \\
0.25  & 1.009843  &106.69  & 1.008633  &108.17  & 1.006354  &109.30  & 1.003194  &110.21  \\
0.30  & 1.011798  &106.76  & 1.010521  &108.23  & 1.008197  &109.36  & 1.005009  &110.27  \\
\hline

\multicolumn{9}{l}{Arginine}\\
0.05  & 1.002629  &121.05  & 1.001683  &122.78  & 0.999571  &124.22  & 0.996520  &125.32  \\
0.10  & 1.005286  &121.13  & 1.004255  &122.90  & 1.002087  &124.30  & 0.999000  &125.40  \\
0.20  & 1.010519  &121.57  & 1.009326  &123.32  & 1.007049  &124.69  & 1.003886  &125.81  \\
0.30  & 1.015742  &121.75  & 1.014396  &123.46  & 1.012010  &124.83  & 1.008776  &125.94  \\
0.40  & 1.020999  &121.76  & 1.019504  &123.44  & 1.017019  &124.77  & 1.013714  &125.88  \\
0.50  & 1.026098  &122.08  & 1.024464  &123.72  & 1.021882  &125.03  & 1.018511  &126.13  \\
\hline

\multicolumn{9}{l}{Tryptophane}\\
0.006 & 1.000369  &132.90  & 0.999498  &134.22  & 0.997434  &136.51  & 0.994413  &137.96  \\
0.008 & 1.000496  &133.97  & 0.999620  &135.64  & 0.997554  &137.70  & 0.994530  &139.30  \\
0.009 & 1.000619  &135.03  & 0.999741  &136.61  & 0.997671  &138.73  & 0.994647  &140.10  \\
0.012 & 1.000151  &189.03  & 0.999275  &190.15  & 0.997207  &191.80  & 0.994178  &193.51  \\
0.015 & 1.000998  &135.96  & 1.000106  &137.92  & 0.998028  &139.88  & 0.994992  &141.69  \\
0.018 & 1.001196  &135.73  & 1.000296  &137.84  & 0.998213  &139.82  & 0.995176  &141.47  \\
0.021 & 1.001035  &152.81  & 1.000135  &154.66  & 0.998050  &156.54  & 0.995011  &158.17  \\
0.024 & 1.001250  &149.89  & 1.000344  &151.79  & 0.998256  &153.60  & 0.995213  &155.26  \\
0.028 & 1.001859  &137.51  & 1.000939  &139.59  & 0.998838  &141.59  & 0.995790  &143.19  \\
\hline

\end{tabular}
\end{table}

\begin{figure}[htb!]
\centerline{\includegraphics[width=0.5\linewidth]{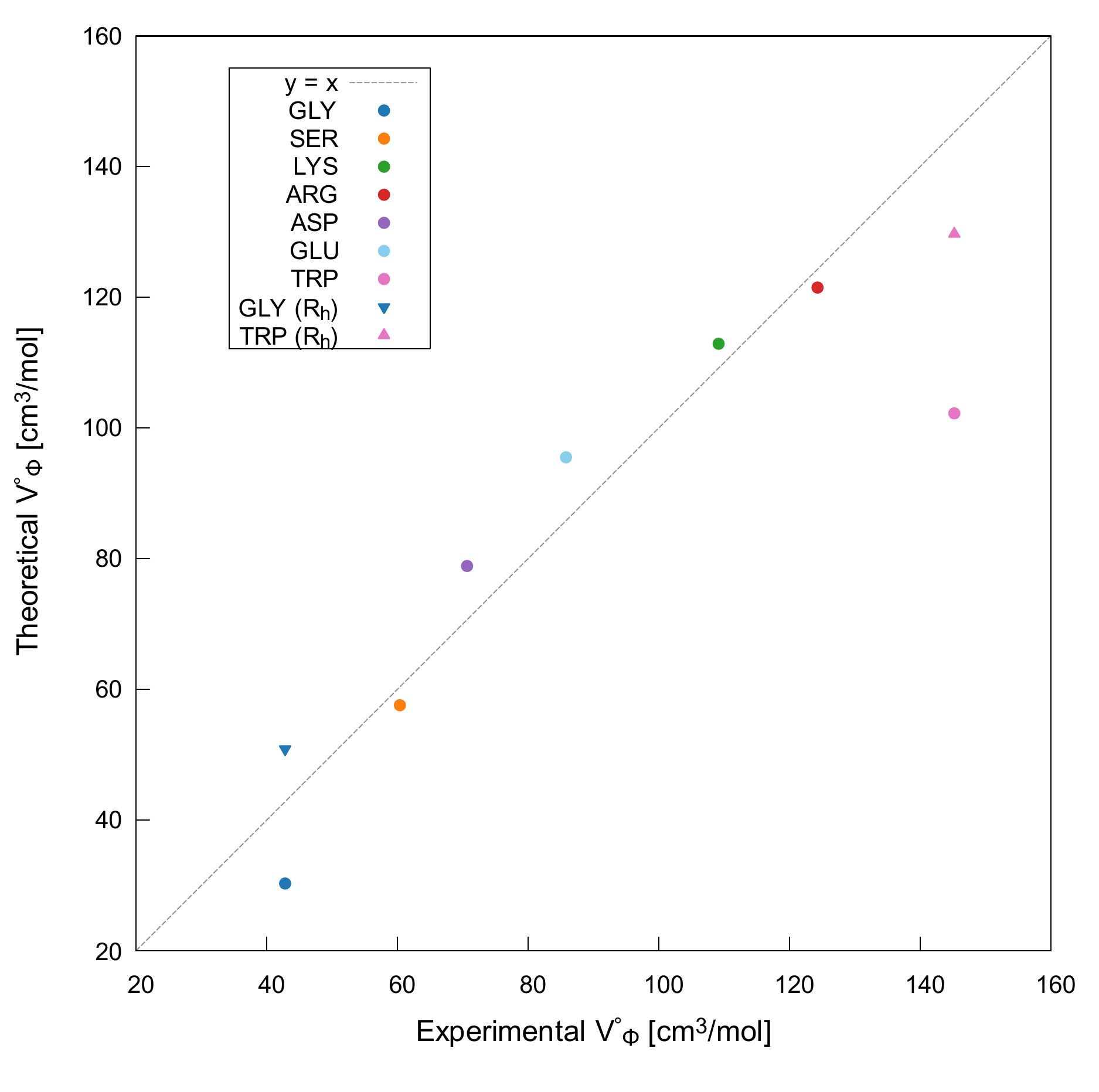}}
\caption{(Colour online) Experimental ($x$-axis) vs.\ theoretically derived ($y$-axis) standard apparent molar volumes
  $V_{\phi}^{\circ}$ for seven $\alpha$-amino acids in water at 298.15 K. Filled circles correspond to the single-bead van der Waals parametrisation; the dotted line indicates perfect agreement ($y = x$). Triangles show results using bead diameters based on hydration radii $2R_h$ for Gly (down triangle) and Trp (up triangle).}
\label{fig:3}
\end{figure}

The KB results for standard apparent molar volumes of water--amino acid mixtures, obtained with the single-bead LJ description and van der Waals based parameters, are presented, together with the experimental results, in figure~\ref{fig:3}. One can see that a relatively good correlation is observed between the two sets of values, and that for the mid-sized residues (Ser, Asp, Glu, Lys, Arg) the theoretical standard molar volumes fall close to the experimental diagonal in $V_{\phi}^{\circ}$ space.

Two amino acids, glycine and tryptophan, deviate more noticeably from this trend. 
In water, glycine carries a particularly tight hydration shell that effectively enlarges the thermodynamic exclusion radius encoded in $G_{12}$ \cite{DAmico13}. We assume that a single van der Waals sphere therefore underestimates $\lvert G_{12}\rvert$, which shifts $V_{\phi}^{\circ}$ to lower values. To check our hypothesis, we repeated the calculation, replacing the van der Waals radius $\sigma$ with the hydration radius $2R_h$ (a solvent-mediated effective size \cite{Germann07}). One can see (figure~\ref{fig:3}) that the newly obtained standard apparent molar volume of glycine coincides, within experimental error, with the experimental $V_{\phi}^{\circ}$.
\begin{table}[htbp]
	\centering
	\caption{Structural first–shell hydration numbers, \(N_{\mathrm{shell}}\), from OZ/HNC for seven \(\alpha\)-amino acids at \(T=298.15\ \mathrm{K}\). The shell cutoff \(r_{\mathrm{cut}}\) is the first minimum of \(g_{12}(r)\) (see equation \ref{eq:N_shell}).}
	\label{tab:5}
	\vspace{0.3cm}
	\begin{tabular}{l c}
		\hline
		Amino acid & \(N_{\mathrm{shell}}\) \\
		\hline
		Gly & 19.3 \\
		Ser & 23.9 \\
		Asp & 27.1 \\
		Glu & 29.0 \\
		Lys & 32.7 \\
		Arg & 34.8 \\
		Trp & 37.2 \\
		\hline
	\end{tabular}
\end{table}

Tryptophan, on the other hand, is bulky and strongly anisotropic due to the indole ring. A single isotropic bead compresses its solvent-accessible area into a sphere and likely underestimates water--aromatic dispersion when cross attractions are set by simple Lorentz--Berthelot mixing. We attempted to improve the result by using $2R_h$ for tryptophan in the theoretical calculation. While this procedure brings the theoretical standard apparent molar volume of tryptophan closer to the experimental value (see figure~\ref{fig:3}), a residual discrepancy still remains, reflecting the neglected anisotropy in the amino acid model.

Overall, the agreement for five amino acids and the diagnostic deviations for Gly and Trp support the use of this Kirkwood--Buff-based route as a transparent bridge between microscopic structure [via $g_{ij}(r)$] and macroscopic volumetric data. The model is intentionally simple, yet it still provides useful and surprisingly accurate estimates of standard molar volumes.

In addition to the experimental volumetric hydration numbers (given in table \ref{tab:2}), we have also calculated the structural hydration number obtained from the amino acid--water pair correlation function (RDF) as:

\begin{equation}
N_{\mathrm{shell}}
= 4\piup\,\rho_1 \int_{0}^{r_{\mathrm{cut}}} g_{12}(r)\, r^{2}\, \mathrm{d}r,
\label{eq:N_shell}
\end{equation}
where \(\rho_1\) is the number density of water molecules and \(r_{\mathrm{cut}}\) is taken as the position of the first minimum in \(g_{12}(r)\) following its first maximum. The results for structural hydration numbers are given in table \ref{tab:5}.

\begin{figure}
    \centering
    \includegraphics[width=0.5\linewidth]{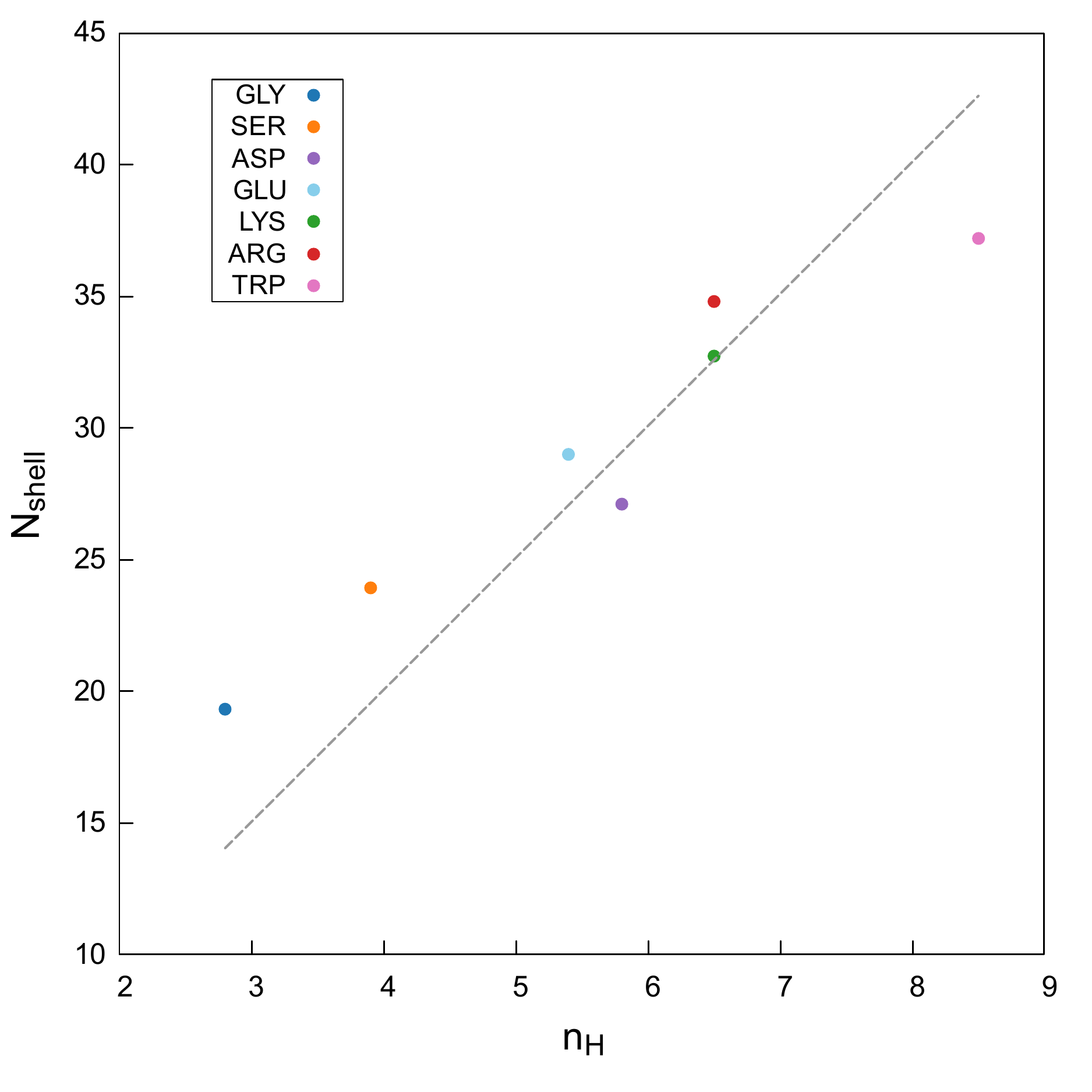}
    \caption{(Colour online) Correlation between the hydration number $n_{\text{H}}$ and the structural hydration number \(N_{\mathrm{shell}}\) for the studied amino acids at \(T=298.15\ \mathrm{K}\). The dashed line represents the linear fit.}
    \label{fig:nH_Nshell}
\end{figure}

The theoretical structural hydration numbers substantially exceed the experimental volumetric numbers. One of the reasons could be that the structural hydration numbers include all the water molecules geometrically located in the first solvation shell of the coarse-grained amino–acid bead (both hydrogen-bonded and non-bonded). By contrast, the experimental volumetric hydration number is a thermodynamic (electrostriction) quantity obtained from volumes and typically amounts to only a few water molecules. Despite the different physical meanings of these measures, $N_{\mathrm{shell}}$ and $n_\text{H}$ exhibit a strong linear correlation across all the studied amino acids (see figure \ref{fig:nH_Nshell}). A linear fit through the origin yields $N_{\mathrm{shell}} = a\,n_\text{H}$ with $a = 5.01$, indicating that both quantities scale consistently with solute size. Reporting both quantities highlights complementary aspects of hydration: local structure (RDF) versus macroscopic response (volumetry).

\section{Conclusion}

In this study, we have examined the applicability of Kirkwood--Buff theory in combination with a simple coarse-grained model to study hydration properties of different types of amino acids. The standard molar volumes were obtained experimentally and compared with the theoretical values. The hydration numbers of the amino acids were calculated, providing valuable insights into solution behaviour and solvation effects. 
The obtained theoretical standard molar volumes are in very good agreement with the experimental values, showing that even a simplified coarse-grained model can yield reliable first estimates for solvation properties. The experimental data obtained were largely consistent with previously reported values, underscoring the reliability of the methodology. We found that the standard molar volumes increase with temperatures, providing insight into temperature effects on electrostriction on water near amino acids. Additionally, the analysis revealed the impact of side-chain contributions on molar volume, offering significant implications for biochemical studies and solution chemistry research. 

\section*{Acknowledgements}
The authors acknowledge financial support from the Slovenian Research and Innovation Agency (research
core funding P1-0201). Z. \v{S}. is grateful for the financial support provided by Slovenian Research and Innovation Agency through the Young Researcher Programme.

\newpage
\ukrainianpart

	\title{Застосування теорії Кірквуда-Баффа для вивчення гідратаційних властивостей $\alpha$-амінокислот}
\author[З. Штефаніч, Б. Грібар-Лі]{З. Штефаніч, Б.~Грібар-Лі}
\address{Факультет хімії та хімічної технології, Університет Любляни, Вечна пот 113, SI-1000 Любляна, Словенія}

\makeukrtitle

\begin{abstract}
	\tolerance=3000%
	Конформаційна стабільність білків, а отже і їхня функціональність, залежить від нековалентних взаємодій, на які сильно впливає навколишнє середовище. Для вивчення властивостей білків, які важко досліджувати безпосередньо, часто використовуються амінокислоти як модельні системи. У цій роботі ми визначили густини семи $\alpha$-амінокислот у водних розчинах у діапазоні температур від 278.15~K до 308.15~K. Згодом ці дані були використані для розрахунку видимих молярних об'ємів. За допомогою лінійної екстраполяції було отримано стандартний молярний об'єм та експериментальні нахили. Потім було проаналізовано об'ємні дані для з'ясування гідратаційної поведінки амінокислот. Внесок бічних ланцюгів у стандартний молярний об'єм було визначено шляхом порівняння властивостей амінокислот з властивостями гліцину. Виявлено, що стандартний молярний об'єм збільшується з температурою внаслідок зменшення електрострикції води навколо молекул амінокислот. Це явище додатково підтверджується зменшенням гідратацiйного числа за вищих температур. Ми використали інтегральне рівняння Орнштейна-Церніке із гіперланцюжковим замиканням (HNC) та грубозернисту модель Леннарда-Джонса (LJ) для розрахунку парних кореляційних функцій. Отримані інтеграли Кірквуда-Баффа для двокомпонентних систем потім були використані для отримання стандартних молярних об'ємів амінокислот. Модель дуже добре відтворює експериментальні стандартні молярні об'єми.	
	\keywords амінокислота, видимий молярний об'єм, стандартний молярний об'єм, гідратаційне число, інтегральне рівняння, теорія Кірквуда-Баффа
\end{abstract}

\end{document}